\documentclass[epjST]{svjour}

\usepackage{amssymb,amsmath}
\usepackage{graphicx,subfigure}
\usepackage{xcolor}

\graphicspath{{./}}

\begin{document}

\title{Active particles in heterogeneous media display new physics}
\subtitle{existence of  optimal noise and absence of bands and long-range order}
\author{Oleksandr Chepizhko\inst{1,2}\fnmsep\thanks{\email{oleksandr.chepizhko@gmail.com}} \and Fernando Peruani\inst{1} \fnmsep\thanks{\email{peruani@unice.fr}}}
\institute{Laboratoire J.A. Dieudonn\'{e}, Universit\'{e} de Nice Sophia Antipolis, UMR 7351 CNRS , Parc Valrose, F-06108 Nice Cedex 02, France \and Department for Theoretical
Physics, Odessa National University, Dvoryanskaya 2, 65026 Odessa,
Ukraine}


\abstract{
We present a detailed study of the large-scale collective properties of self-propelled particles (SPPs) moving in two-dimensional heterogeneous space. 
The impact of spatial heterogeneities on the ordered, collectively moving phase is investigated. 
We show that for strong enough spatial heterogeneity, the well-documented high-density, high-ordered propagating bands that emerge in homogeneous space disappear. 
Moreover, the ordered phase does not exhibit long-range order, as occurs in homogeneous systems, but rather quasi-long range order: i.e. the SPP system becomes disordered in the thermodynamical limit. 
For finite size systems, we find that there is an optimal noise value that maximizes order. Interestingly, the system becomes disordered in two limits, for high noise values as well as for vanishing noise. This remarkable finding  strongly suggests the existence of two critical points, instead of only one, associated to the collective motion transition. 
Density fluctuations are consistent with these observations, being higher and anomalously strong at the optimal noise, and decreasing and crossing over to normal for high and low noise values. Collective properties are investigated in static as well as dynamic heterogeneous environments, and by changing the symmetry of the velocity alignment mechanism of the SPPs. 
} 


\maketitle

\section{Introduction}
\label{intro}

We understand by an {\it active particle}, a particle that is able to convert energy into work to self-propel in a dissipative medium. 
There are several strategies to achieve  self-propulsion.   
(i) Particles may possess an energy depot and be equipped with a motor~\footnote{Motility assays represent a particular case, where motors are not attached to the moving particles, while the energy depot could be considered as extended over the space~\cite{schaller2010}}, a scenario representative of  many biological systems such as  
moving bacteria~\cite{zhang2010,peruani2012}, insects~\cite{Buhl_science_locust,romanczuk2009}, and animals~\cite{Huepe_fish_2013,Holdo2011}.  
(ii) Particles may be able to rectify an external driving in order to achieve self-propulsion in a given direction, 
a situation commonly observed in non-living, artificial active particles such as vibration-induced self-propelled rods and discs~\cite{deseigne2010,kudrolli2008,weber2013}, 
light-induced thermophoretic  active particles~\cite{jiang2010,golestanian2012,theurkauff2012,palacci2013}, chemically driven particles~\cite{golestanian2009,paxton2004,mano2005,rucker2007,howse2007,golestanian2005}, 
and rollers driven by the Quicke rotation effect~\cite{spp_activecolloid_nature2013}. 
Independently of the strategy exploited by the particles to self-propel, these systems are intrinsically out of (thermodynamic) equilibrium~\footnote{This is due to the  energy consumption involved by the self-propelling mechanism and the energy dissipated to the medium}  and even 
 in absence of particle-particle interactions exhibit a non-trivial behavior. For instance, fluctuations in the self-propelling mechanism can lead to complex transients in the mean-square displacement of the particles~\cite{peruani2007} as well as anomalous velocity distributions~\cite{romanczuk2012}. 

Interestingly, most, if not all, active particle systems found in nature take place, at all scales, in the heterogeneous media:  
%
from bacterial motion in natural habitats~\cite{dworkin}, such as the gastrointestinal tract and the soil, among other complex environments, to the migration of herd of mammals across forests and steppes~\cite{Holdo2011}. 
Despite this evident fact, active matter research has focused almost exclusively, at the experimental and theoretical level, on homogeneous active systems~\cite{marchetti2013,ramaswamy2010,vicsek2012,romanczuk2012,ramaswamy2003}.   
Non-equilibrium, large-scale properties of active systems such as long-range order in two-dimensions as Vicsek et al.~\cite{vicsek1995} reported in their pioneering paper,  the 
emergence of high-order, high-density traveling bands~\cite{gregoire2004,caussin2014}, and the presence of giant number fluctuations in ordered phases~\cite{ramaswamy2010,spp_tonertu_prl1995,toner1998} are all non-equilibrium features either predicted or discovered in perfectly homogeneous systems. 
Here we show that most of these non-equilibrium features are strongly affected by the presence of spatial heterogeneities. Moreover, we show that these properties vanish in strongly  heterogeneous media. 
More specifically, we extend previous results~\cite{ChepizhkoAltmannPeruani2013} on the large-scale collective properties of interacting self-propelled particles (SPPs) moving at constant speed in an  heterogeneous space.  
We model the spatial heterogeneity as a random distribution of undesirable areas or ``obstacles"  that the SPPs avoid. The degree of heterogeneity is controlled by the average density $\rho_o$  of obstacles. 
We provide numerical evidence that indicates that at  low densities of heterogeneities $\rho_o$, the SPPs exhibit, below a critical noise intensity $\eta_{c1}$, long-range order (LRO). 
For noise intensities $\eta$ close to $\eta_{c1}$, the SPPs self-organize, as in homogeneous space, into high-density traveling structures called ``bands". 
%
We find that  as $\rho_o$ is increased, bands become less pronounced   
to the point that for  large enough values of $\rho_o$ they are no longer observed.
Our results indicate that in strongly heterogeneous media, i.e. large values of $\rho_o$, the large-scale  properties of the system are remarkably different from what we know of the Vicsek model~\cite{vicsek1995} in two-dimensional homogeneous media.   
For instance, orientational order for $\eta<\eta_{c1}$ is no longer LRO, but rather quasi-long range (QLRO), with the system exhibiting the maximum degree of order 
at an intermediate noise value $\eta_M$, such that $0<\eta_M<\eta_{c1}$. Moreover, we provide solid evidence  that the system becomes disordered as $\eta \to 0$. 
The numerical data suggests the existence of a second critical point $\eta_{c2}$, with $0<\eta_{c2}<\eta_{c1}$  below which the system is genuinely disordered. The disordered phase at low $\eta$ values is characterized by the presence of large, dense moving clusters.  
We show that the particle number statistics is consistent with these observations: giant number fluctuations (GNF) are high~\footnote{The associated GNF exponent adopts its maximum value near $\eta_M$.} near $\eta_M$ and decrease as $\eta$ approaches $\eta_{c2}$, with GNF becoming weaker as $\rho_o$ is increased to the point that fluctuations become normal.  
Finally, we investigate and compare static and dynamical heterogeneous media, as well as SPPs with ferromagnetic and nematic velocity alignment. We show that in all cases there exists  an optimal noise intensity that 
maximizes the ordering in the systems. 
The reported results might be of a great importance for the design and control of active particles systems.

The paper is organized as follows. In Section~\ref{sec:model} we introduce a general and simple model for SPPs in heterogeneous media. The collective motion phase, the associated kinetic phase transition and ordering properties of the model are studied in Sec.~\ref{sec:order}. 
Anomalous density fluctuations are addressed in Sec.~\ref{sec:gnf}. 
In section~\ref{sec:discussion}, we investigate the collective properties of SPPs in a dynamical heterogeneous environment, and discuss  the effect of the velocity alignment symmetry.  
We summarize our results in Sec.~\ref{sec:conclusion}.

\section{Formulation of the model}
\label{sec:model}

\begin{figure}
\centering
\includegraphics[scale=0.35]{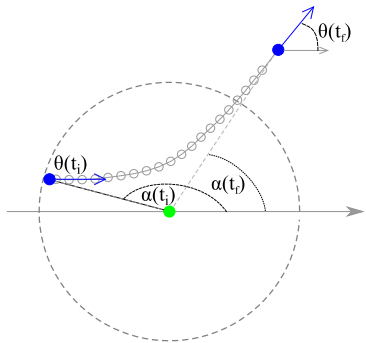}
\caption{Interaction of a self-propelled particle (blue) with an obstacle (green). Angles are given with respect to horizontal axis that is directed from left to right. The moving direction given by the angle $\theta(t)$ evolves in time from an  initial direction at $t_i$ to a final direction at $t_f$  final. The temporal evolution of the particle position, with respect to the obstacles, is given by the angle $\alpha(t)$. The initial value of $\alpha$ at $t_i$ and its final value at $t_f$ are marked with dashed lines. Notice, that after a collision, $\theta(t_f) \approx \alpha(t_f)$, the difference $\Delta = \theta(t_f) - \alpha(t_f)$ is such that $\Delta << 1$ for either large values of $\gamma_o$ or large values of $R_o$. In the figure, $\gamma_o = 1$ and $R_o=1$.}.
\label{fig:interaction}
\end{figure}

We consider a system of $N$ self-propelled particles, moving with a constant speed $v_0$ on a two-dimensional heterogeneous space of linear size $L$ with periodic boundary conditions. 
We express the equations of motion of the $i$-th particle as:
\begin{eqnarray}
	\dot{\mathbf{x}}_i &=& v_o \mathbf{V} (\theta_i)
\label{eq:position} \\
	\dot{\theta}_i &=& g(\mathbf{x}_i) 
	\left[
		\frac{\gamma_b}{n_b (\mathbf{x}_i)}
		\sum_{|\mathbf{x}_i - \mathbf{x}_j| < R_b}
		\sin\left[ q(\theta_j - \theta_i) \right]
	\right]
	+
	h(\mathbf{x}_i, \theta_i) + \eta \xi_i (t) \,,
\label{eq:direction}
\end{eqnarray}
where the dot denotes temporal derivative, $\mathbf{x}_i$ corresponds to the position of the $i$th particle, and $\theta_i$ is an angle associated to its moving direction.  
Equation~(\ref{eq:position}) indicates simply that the particle moves at speed $v_0$  in direction $\mathbf{V}(\theta_i)=(\cos\theta_i,\sin\theta_i)$. 
Equation~(\ref{eq:direction}) conveys the dynamics of the moving direction of the particle, which is parametrized by the angle $\theta_i$. The first term on the right-hand side corresponds 
to a velocity-velocity alignment mechanism acting between neighboring particles as in the  Vicsek model~\cite{vicsek1995}, 
the second term models the interaction of the $i$-th particle with the spatial heterogeneities, and the third term 
is an additive noise, where $\langle \xi_i(t) \rangle=0$, $\langle \xi_i(t) \xi_j(t') \rangle=\delta_{ij}\delta(t-t')$, and $\eta$ denotes the noise strength. 
The velocity-velocity alignment is characterized by three parameters: its symmetry, given by $q$ and that we fix for most of this analysis to be ferromagnetic with $q=1$, the interaction radius $R_b$, and the alignment strength $\gamma_b$, where the symbol $n_b(\mathbf{x}_i)$ represents 
the number of SPPs at a distance less than or equal to $R_b$ from $\mathbf{x}_i$, i.e. the number of neighbors of the $i$-th particle. 
The function $g(\mathbf{x}_i)$ controls the relative weight between alignment (to the other particles) and heterogeneity/obstacle avoidance. 
We tested two possibilities, both leading to the same macroscopic behavior: $g(\mathbf{x}_i)=1$ and  $g(\mathbf{x}_i)=1-\Theta(n_o(\mathbf{x}_i))$,
 where $\Theta(x)$ is  Heaviside step function. The latter implies that in the proximity of an obstacle, the SP particle focuses on avoiding it, without aligning to the neighbors. 
Results shown here correspond to this definition. 
Finally, the interaction with the spatial heterogeneities, which we refer to as ``obstacles'' or undesirable areas, is given by the function $h(\mathbf{x}_i, \theta_i)$, defined as:
\begin{eqnarray}
	h(\mathbf{x}_i, \theta_i) = \frac{\gamma_o}{n_o(\mathbf{x}_i)}
	\sum_{|\mathbf{x}_i - \mathbf{y}_k| < R_o}
	\sin(\alpha_{k, i} - \theta_i) \, ,
	\label{eq:for_h} 
\end{eqnarray}
if $n_o(\mathbf{x}_i)>0$, otherwise $h(\mathbf{x}_i) = 0$. The term $n_o(\mathbf{x}_i)$ represents the number obstacles at a distance less than or equal to $R_o$. 
The angle $\alpha_{k, i}$ is simply the angle of the polar representation of the vector $\mathbf{x}_i - \mathbf{y}_k = \Gamma_{k, i}\, (\cos \alpha_{k, i}, \sin \alpha_{k, i})$, where $\mathbf{y}_k$ is the position of the $k$th-obstacle and $\Gamma_{k, i}$ is the norm of the vector. The position $\mathbf{y}_k$ of obstacles, with $k \in [1, N_o]$, is random and homogeneously distributed in space.   
The interaction between a SPP and an obstacle is depicted in figure~\ref{fig:interaction}: as the particle approaches the obstacle, its trajectory is deflected. 
While here we focus mainly on ``obstacles" whose position is fixed in time, we will also address briefly, at the end of the paper, the case of moving obstacle, i.e. free to diffuse around the space. 

\begin{figure}
\centering
\includegraphics[scale=0.4]{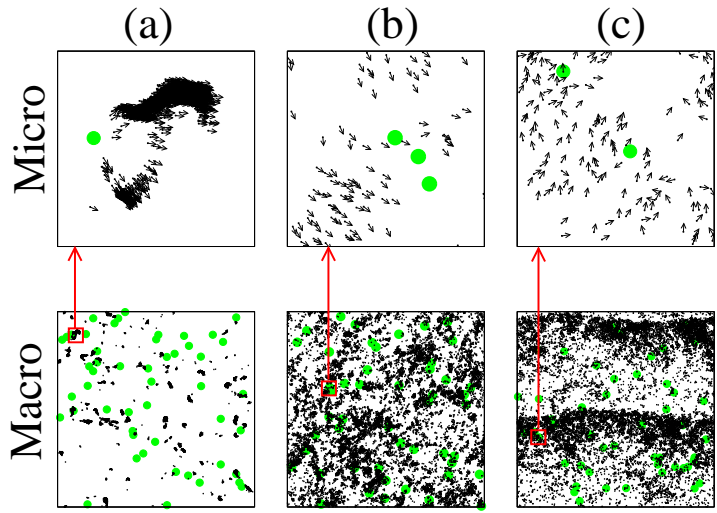}
\caption{ 
Simulation snapshots of  different phases for $\rho_o = 2.55 \times10^{-3}$ for a system size $N_b=19600$ ($L=140$). The SPP are represented as black arrows, while the obstacles as green dots. The bottom panels correspond to macroscopic phases, while the top panels to zoom up regions, where the interaction between the SPPs and the obstacles can be appreciated. From left to right:  (a) clustered phase, $\eta=0.01$ and $r=0.58$, (b) homogeneous ordered phase, $\eta=0.3$ and $r=0.97$, and (c) ordered band phase, $\eta=0.6$ and $r=0.73$.}
\label{fig:snapshots}
\end{figure}

It is worth analyzing few simple limits of the model given by equations~(\ref{eq:position}) and~(\ref{eq:direction}). 
For $\gamma_b=\gamma_o=0$, the equations define a system of non-interacting persistent random walkers characterized by a diffusion coefficient $D_x = v_o^2/\eta^2$. 
With $\gamma_o=0$ and $\gamma_b>0$ (or equivalently $\gamma_o>0$ and $N_o=0$), the model reduces to a continuous-time version~\cite{peruani2008} of the Vicsek model~\cite{vicsek1995}. 
For $\gamma_o>0$, $\gamma_b=0$, and $N_o>0$, the equations describe a system of non-interacting active particles moving at constant speed on an heterogeneous space, where several interesting non-equilibrium features can be observed~\cite{ChepizhkoPeruani2013}. 
At low obstacle density, particles move diffusively with a  diffusion coefficient that is, interestingly, a non-monotonic function of the obstacle density. It  
reaches a minimum at a given, non-trivial, non-zero obstacle density. 
On the other hand, at high obstacle density and for large enough interaction strength $\gamma_b$, spontaneous trapping of particles can occur. In this scenario,  particles move sub-diffusively across the space, spending arbitrary long time in the spontaneously formed traps~\cite{ChepizhkoPeruani2013}. 
Here, we focus on the general scenario where $\gamma_o>0$, $\gamma_b>0$, and $N_o>0$. 
We reduce the parameter space by fixing the following parameters:  $R_b=R_o=1$, $\gamma_b=\gamma_o=1$, $\rho_b=N_b/L^2=1$, $v_0=1$ and a discretization time step to $\Delta t=0.1$. 
Notice that our main control parameters are the noise intensity $\eta$ and the number of obstacles $N_o$ or equivalently, the obstacles density $\rho_o=N_o/L^2$. 

\begin{figure}
\centering
\includegraphics[scale=0.3]{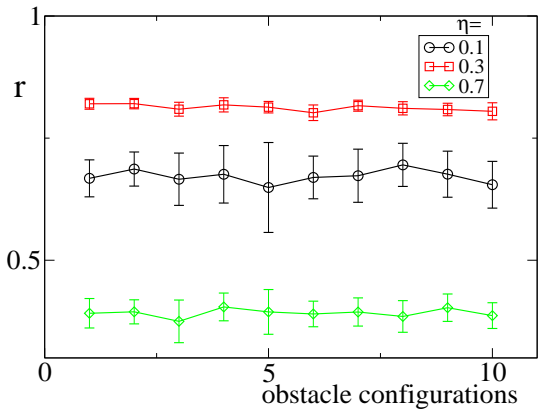}
\caption{Average values of order parameter $r$ for 10 different initial configuration of obstacles for three values of the noise strength $\eta$ and $\rho_o=0.0325$. System size $N_b=10000$. Notice  that the average value of the order parameter does not depend on the particular configuration of the randomly placed (static) obstacles. The error bars correspond to the standard deviation of the time series of $r$ obtained for each obstacle configuration and set of parameters.}
\label{fig:ten_realizations}
\end{figure}

\section{The order-disorder transition}
\label{sec:order}

The system exhibits three distinct macroscopic phases, for any given obstacle density $\rho_o>0$, as we move from high to low noise amplitude $\eta$. 
At high noise values, particles are homogeneously distributed in space as a disordered gas of non-interacting particles. 
Below a critical noise value $\eta_{c1}$, the system undergoes a kinetic phase transition from the disordered gas to a (locally) ordered phase. 
For finite size systems, there is a symmetry breaking and the ordered phase implies the existence of a net flux of particles in a given direction, or equivalently the existence of a preferred direction of motion. 
For values close to onset of collective motion, i.e. close to $\eta_{c1}$, and for rather low values of $\rho_o$ particles self-organize into high-density, high-order traveling structures called ``bands", as illustrated in Fig.~\ref{fig:snapshots}(c). 
As $\eta$ is decreased further, getting deeper in the ordered phase, bands disappear and we observe an ordered phase where particles are roughly homogeneously distributed in space,  though anomalously large density fluctuations are present, Fig.~\ref{fig:snapshots}(b). 
If  $\eta$ is decreased even further, coming close to the noiseless limit, counterintuitively the degree of order drops dramatically and particles are organized into densely packed moving clusters, see Fig.~\ref{fig:snapshots}(a),  that are only weakly correlated due to the constant deflections they experience when running into obstacles. 

\begin{figure}
\centering
\resizebox{0.9\columnwidth}{!}{%
  \includegraphics{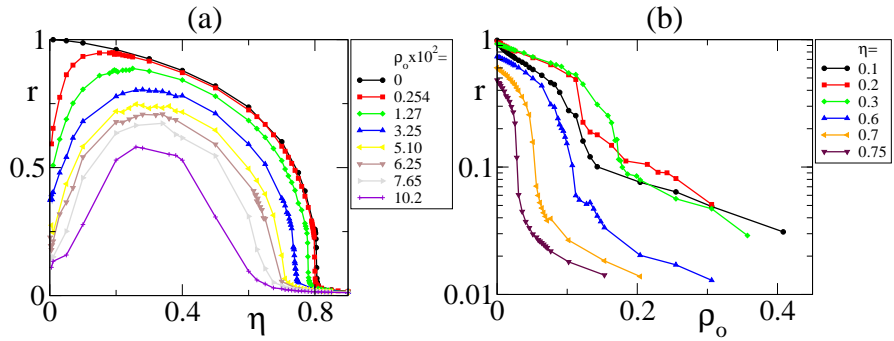} }
\caption{ (a) Order parameter $r$ vs.  noise strength $\eta$ for various values of the obstacle density  $\rho_o$. (b) Order parameter $r$ vs. obstacle density  $\rho_o$ for various values of noise strength $\eta$. 
 System size: $N_b=19600$ ($L=140$). Notice in (a) that curves for $\rho_o>0$ exhibit a (local) maximum. This implies the existence of optimal noise value that maximizes collective motion.}
\label{fig:order_parameter}       
\end{figure}

In order to quantify the degree of order in the system, we use the following order parameter:
\begin{equation}
	r =
	\langle	r(t) \rangle_t =
	\left\langle
		\left|
		\frac{1}{N_b}
		\sum_{i=1}^{N_b}
		e^{i \theta_i(t)}
		\right|
	\right\rangle_t\,,
	\label{eq:definition_of_order_parameter}
\end{equation}
where $\langle \dots \rangle_t$ denotes the average over time~\footnote{Since our initial condition is typically a random distribution of particles over space with random moving direction, this average is performed after removing the transient.}. 
The definition of $r$ is simply, in complex notation, the norm of the average velocity of the particles. 
Values of $r$ larger than zero indicate that there is a net flux of particles in a given direction, while $r=0$ corresponds to a disordered system. 
It is important to stress that the average value of $r$ does not depend on the particular configuration of (static) obstacles, as long as the obstacles have been placed randomly in the space. 
For instance, if we compare simulations performed with different obstacle configurations, we obtain the same average value of  $r$, as shown in Fig.~\ref{fig:ten_realizations}, where the error bars represent 
the measured standard deviation in the corresponding time series of the order parameter $r$.  
In summary, the value of $r$, Eq.~\ref{eq:definition_of_order_parameter}, depends only on the noise amplitude $\eta$ and obstacle density $\rho_o$.

Fig.~\ref{fig:order_parameter} shows the dependency of the order parameter $r$ with respect to the noise intensity $\eta$ for various obstacle densities $\rho_o$,  panel (a), 
and with respect to the obstacle density for various noise intensities, panel (b). 
The curve that corresponds to $\rho_o=0$ in Fig.~\ref{fig:order_parameter}(a), black curve, exhibits the behavior that we expect in an homogeneous space, i.e. as expected in the classical formulation of the Vicsek model~\cite{vicsek2012}. 
This reference curve indicates that in an homogeneous system, the maximum order is reached in the noiseless limit. 
Notice that for an homogeneous system, as the noise intensity is increased, the order parameter $r$ monotonically decreases until 
 the critical noise strength $\eta_{c1}$,  above which the system is fully disordered and $r=0$. 
Fig.~\ref{fig:order_parameter}(a) shows that 
the presence of even a small amount of  obstacles leads to a qualitatively different picture, with the order parameter $r$ exhibiting a different behavior with $\eta$.  
All curves with $\rho_o>0$ reach a maximum at a non-zero value of $\eta$, and all decrease as the noiseless limit is approached. 
This means that for each value of  $\rho_o$, there is an optimal value of $\eta$ that maximizes collective motion. We refer to this $\eta$ value as  $\eta_M$. 
On the other hand, we observe that if we fix the noise intensity $\eta$ and vary the obstacle density $\rho_o$, as displayed in Fig.~\ref{fig:order_parameter}(b), the order parameter $r$ 
monotonically decreases as $\rho_o$ is increased. Notice that curves corresponding to different noises exhibit a quite distinct behavior, for instance compare the curve for $\eta=0.3$, close to the optimal noise $\eta_M$ for most $\rho_o$ values, with the other curves.

In the following we divide our quantitative analysis into two statical data sets that correspond to  low obstacle density and high obstacle density, respectively. We show that at quite small obstacle densities, the numerical data 
is consistent with a discontinuous (kinetic) phase transition. The numerical data indicates that as the obstacle density increases, the traveling bands become weaker until they disappear. We show that once we reach such obstacle densities (i.e. for $\rho_o\geq0.1$), the order is no longer long-range (LRO) in the ordered phase, but rather quasi-long range (QLRO). Our results unambiguously indicate that at such high obstacle densities, as we approach the noiseless limit, 
i.e. when particles self-organize into densely packed moving clusters as shown in Fig.~\ref{fig:snapshots}(a), the system is fully disordered, which suggests the existence of a second critical point. 

\begin{figure}
\centering
\includegraphics[scale=0.42]{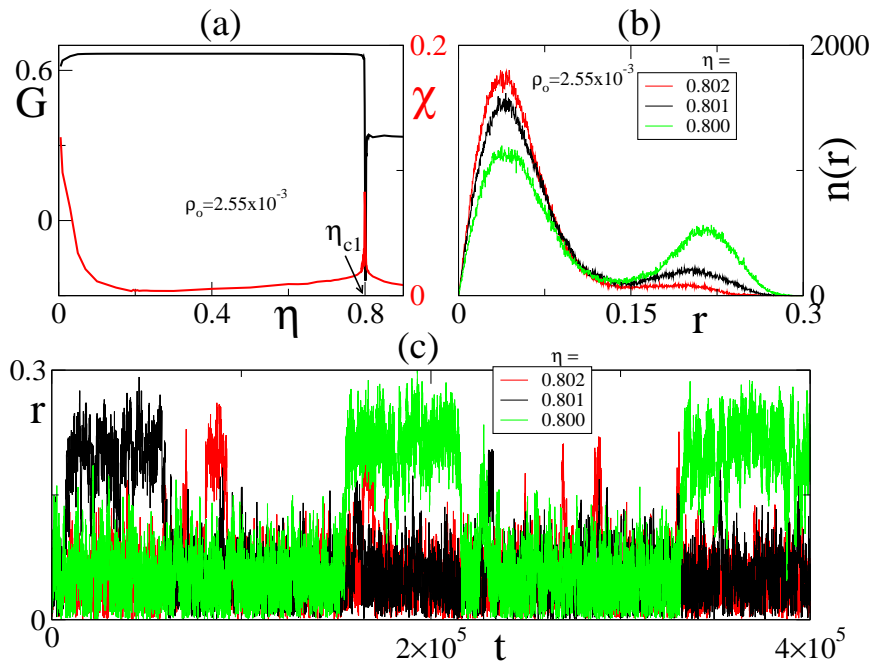}
\caption{Evidence of a first-order phase transition at low obstacle densities. 
(a) Binder cumulant $G$ and susceptibility $\chi$ as function of the noise intensity $\eta$. Notice that around $\eta\sim 0.8$, $G$ reaches negative values and $\chi$ peaks.  
(b) Histogram $n(r)$ of the order parameter obtained from time series of $r$. (c) Time series of the order parameter $r$. Notice  the flip-flops in the  time series. System size $N_b=19600$ ($L=140$).}
\label{fig:first_order_evidence}
\end{figure}

\subsection{Low obstacle density}

To characterize the phase transition to orientational order at low obstacle densities, we introduce two additional quantities, the susceptibility $\chi$ and the Binder cumulant~\cite{Binder} $G$, whose definitions are given by:
\begin{eqnarray}
	\chi = \langle r^2 \rangle_t - \langle r \rangle_t^2\, ,
	\label{eq:variance_definition}  \\
	G =
	1 - \frac{\langle r^4 \rangle_t}{3 \langle r^2 \rangle_t^2}\,,
	\label{eq:Binder_definition} \, .	
\end{eqnarray}
We use $\chi$ to determine precisely the position of the critical point $\eta_{c1}$ and $G$ to estimate whether the probability distribution of the order parameter $r$ is unimodal and Gaussian. Notice that the definition of $G$ is directly related to the excess kurtosis. 
Fig~\ref{fig:first_order_evidence}(a) shows both quantities as function of the strength $\eta$ for one of the smallest obstacle densities tested,  $\rho_o=2.54\times 10^{-3}$. 
The peak of $\chi$ and the sudden change of behavior of $G$ at $\eta=0.8$ indicates that  the critical point is located at $\eta_{c1}=0.8$. 
The drop of $G$ to negative values suggests that the probability distribution of $r$ is bimodal~\cite{chate2008}, as confirmed in Fig.~\ref{fig:first_order_evidence}(b). 
This finding is the result of abrupt transitions in the value of $r$ along time, often called ``flip-flops", from low values (disordered gas phase) to high values (ordered phase) and vice versa, Fig.~\ref{fig:first_order_evidence}(c). 
%
%
In summary, the statical data close to the critical point $\eta_{c1}$ is consistent with a discontinuous (kinetic) phase transition: negative values of $G$, a bimodal distribution, and flip-flops as $\eta \to \eta_{c1}$.  

\begin{figure}
\centering
\includegraphics[scale=0.3]{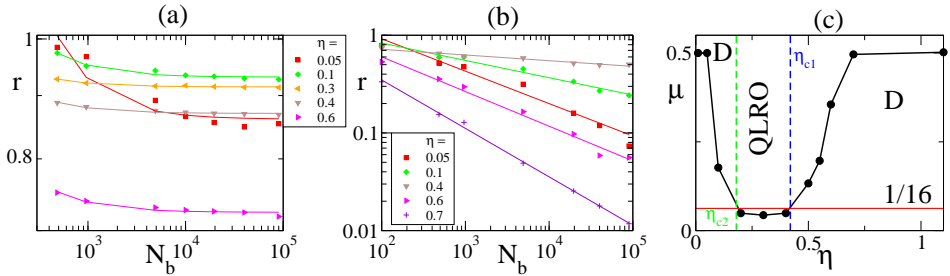}
\caption{ Finite size analysis at low and high obstacle densities. 
(a) Scaling of $r$ as function of the system size $N_b$ for low obstacle densities, here $\rho_o = 2.55 \times 10^{-3}$, for various values of $\eta$ (color coded). The solid lines correspond to exponential fittings: $r(N_b) = r_{\infty} + C_* \exp(-N_b/N_*)$, with $r_{\infty}$, $C_*$, and $N_*$ fitting constants. Notice that for $N_b \to \infty$, the order parameter reaches a constant value $r(N_b\to \infty) \to r_{\infty}$, i.e. long-range order (LRO). 
(b) Scaling of $r$ as function of the system size $N_b$ for high obstacle densities, here $\rho=0.102$,  for various values of $\eta$ (color coded). The solid curves correspond  to power-law fittings: $r(N_b) = A \, N_b^{-\mu}$, where $A$ and $\mu$ are fitting constants. We define, see text, quasi-long-range order (QLRO) when $\mu<1/16$. On the other hand, $\mu=1/2$ implies that the system is fully disordered. (c) Exponent $\mu$ as function of the noise intensity $\eta$ for $\rho=0.102$. The diagram allows us to define two critical points, indicated by the vertical lines: the vertical line to the left corresponds to $\eta_{c2}$, while the other one to $\eta_{c1}$. In between $\eta_{c2}<\eta<\eta_{c1}$ the order is QLRO. The value $1/16$ is indicated by an horizontal red line.}
\label{fig:finite_size_scaling}
\end{figure}

Our next step is to determine whether at low obstacle densities the observed order for $\eta<\eta_{c1}$  remains present in the thermodynamical limit. 
This involves a finite size scaling. The goal is to obtain the scaling of the order parameter $r$ with the system size. If the system exhibits  long-range order (LRO), by increasing the system size, while keeping 
constant the particle density $\rho_b$ and obstacle density $\rho_o$, we expect $r$ to saturate to a non-zero value.  
As measure of the system size we use $N_b$~\footnote{As measure of system size, instead of $N_b$, we can use either $N_o$ or $L$.  Since $\rho_b$ and $\rho_o$ are constant, knowing either $L$, $N_b$, or $N_o$, we can determine the other two.}. 
Fig.~\ref{fig:finite_size_scaling}(a) shows that at low obstacles densities, here $\rho_o=2.55\times 10^{-3}$,  $r$ effectively saturates with $N_b$ to a non-zero value for a large range of $\eta<\eta_{c1}$ values~\footnote{Unfortunately, we can not be sure that this behavior  remains for $\eta<0.05$. Smaller noises are hard to explore numerically.}.  
Thus, the numerical data at very low densities, up to the system size we could reach, is consistent with LRO. This implies that in the thermodynamical limit we expect the system to remain ordered, i.e. 
as $N_b \to \infty$, $r\to r_{\infty}(\eta)>0$, where $r_{\infty}(\eta)$ is the asymptotic value of $r$ in an infinite system, which is a function of $\eta$, $\rho_b$, and $\rho_o$.

\begin{figure}
\centering
\includegraphics[scale=0.35]{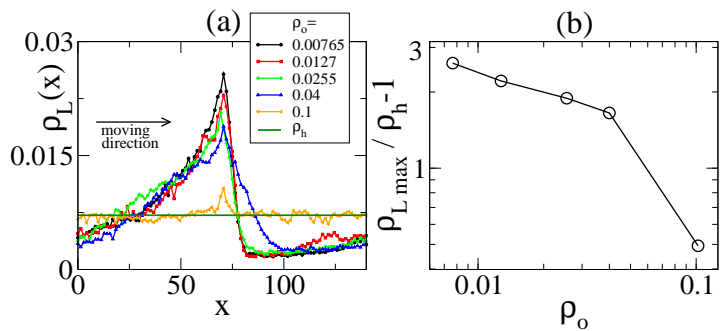}
\caption{Bands.  (a) The particle density profile $\rho_L(x)$ of the bands along the (band) moving direction. 
The horizontal line indicates the density profile corresponding to an homogeneous distribution of particles whose value we denote by $\rho_h$. 
(b) Maximum $\rho_{L\, max}$ of the density profile $\rho(x)_L$ vs. the density of obstacles. 
The panel shows the quantity $\rho_{L\, max}/\rho_H -1$, which drops to zero for large densities, which indicates the absence of bands in the system. 
}
\label{fig:bands}
\end{figure}

At low obstacle densities, as mentioned above, we observe close to the critical point $\eta_{c1}$ a behavior consistent with a first-order (discontinuous) phase transition.  
This seems to be related~\cite{gregoire2004} with the emergence of  high-order traveling bands~\footnote{For instance, when flip-flops are observed, high values of $r$ coincide with the emergence of bands. Along the simulation, bands appear and disappear constantly, as flip-flips do.}, as shown in  Fig.~\ref{fig:snapshots}(c). 
Bands are narrow structures that expand through the whole system, elongated in the direction perpendicular to their direction of motion.
Several density profiles, corresponding to various $\rho_o$ values, are displayed in Fig.~\ref{fig:bands}(a). To construct these profiles, one needs to project the particle positions onto the moving direction of the band and  make a histogram. 
Bands exhibit a sharp front and a smooth tail. They move across a background gas of disordered (or weakly ordered) particles and accumulate particles in the front as they advance forward, while loosing particles in the rear.  
The presence of obstacles strongly affects bands. As $\rho_o$ is increased, profiles get smoother as illustrated in Fig.~\ref{fig:bands}(a). 
This can be more quantitatively seen in Fig.~\ref{fig:bands}(b) that shows how the bands vanish for large value of $\rho_o$.
In short, as the number of obstacles is increased, bands become weaker, with a profile that  relaxes towards the background gas. At some point, bands and the background gas are undistinguishable and bands vanish. This is evident for $\rho_o>0.1$, where bands are no longer observed.

\subsection{High obstacle density}
 
At high enough obstacle densities, where bands are no longer observed, the system behavior is remarkably different. 
The finite size study reveals that the system is unable to reach LRO at such high $\rho_o$ values. 
We find that the order parameter $r$ decays with the system size $N_b$ as a power-law: $r \propto N_b^{-\mu(\eta)}$, where the exponent $\mu(\eta)$ is a function of $\eta$, as shown in Fig.~\ref{fig:finite_size_scaling}(b) for $\rho_o=0.102$. 
For  noises close to the optimal value $\eta_M$,  we obtain exponent values such that $0 < \mu < 1/16$. Here, by analogy with Kosterlitz-Thouless transition~\cite{Kosterlitz} we say that when $0 < \mu < 1/16$ there is quasi-long range order (QLRO), while for $ \mu>1/16$ we assume that the system is disordered. 
Fig.~\ref{fig:finite_size_scaling}(c) shows that the behavior of $\mu(\eta)$ is such that we can define two disordered phases, one at high noise values and the other one at low noises, and the ordered phase
with QLRO at intermediate noises. 
This implies the existence of two critical points, which obey $\mu(\eta_{ci})=1/16$, with $i=1, 2$ such that $\eta_{c2}<\eta_{c1}$, see  Fig.~\ref{fig:finite_size_scaling}(c). 
The system exhibits QLRO when $\eta_{c2}<\eta<\eta_{c1}$, while being disordered for $\eta<\eta_{c2}$ and $\eta>\eta_{c1}$. 
Notice that for both disordered phases, $\mu$ reaches $1/2$, in particular we observe that $\mu \to 1/2$ in two limits:  $\eta \to 0$ and $\eta \to \infty$. 
A scaling  $r \propto N_b^{-1/2}$ corresponds to a fully disordered system with a random distribution of moving directions. 
Interestingly, the densely packed moving cluster phase at low $\eta$ values, as the one observed in Fig.~\ref{fig:snapshots}(a), corresponds, at high obstacle densities (i.e. for $\rho_o\geq 1$), to a fully disordered phase. 
The presence of only QLRO, or rather the absence of LRO, implies that in the thermodynamical limit we expect $r \to 0$ for all $\eta>0$ values, i.e. we expect an infinite system to be disordered. 
Nevertheless, between $\eta_{c2}<\eta<\eta_{c1}$, we expect the SPPs display large correlations in their moving direction also in the thermodynamical limit.

%
%
%
%

%

%
\begin{figure}
\centering
\includegraphics[scale=0.35]{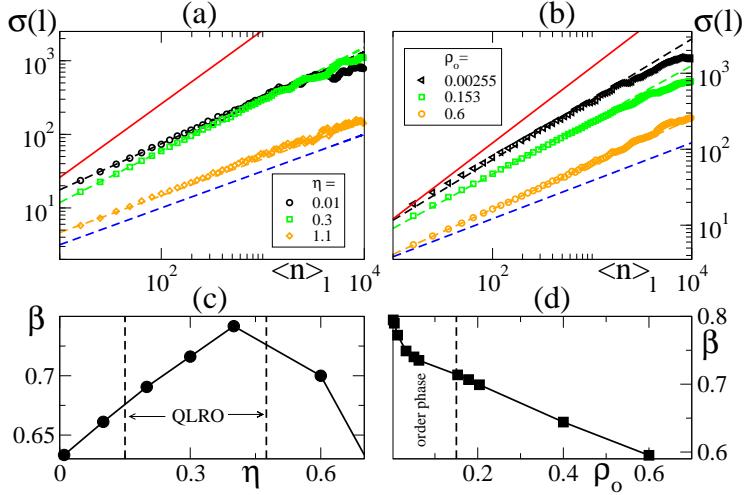}
\caption{
Number fluctuation statistics. 
(a) Scaling of the variance of the particle number $\sigma(l)$ vs. the average particle number $\langle n \rangle_l$ for different values of noise strength $\eta$ for $\rho_o= 0.102$. 
(b) Scaling of $\sigma(l)$ vs $\langle n \rangle_l$ for different values of the density of obstacles $\rho_o$ and fixed noise strength $\eta=0.3$. 
In (a) and (b), the scaling $\sigma(l) \propto \langle n \rangle_l^1$ and $\sigma(l) \propto \langle n \rangle_l^{1/2}$ - used as reference - are indicated by a solid red and a dashed blue curve, respectively. The dashed lines in (a) and (b) represent power-law fitting curves. 
(c) The dependency of the scaling exponent $\beta$ (see text for more details) on the noise strength $\eta$ at fixed $\rho_o = 0.102$. 
For $\rho_o = 0.102$, $\eta_{c1} \approx 0.45 \pm 0.05$ and $\eta_{c2} \approx 0.15 \pm 0.05$, indicated by two dashed vertical lines in the figure. In the range $\eta_{c2}<\eta<\eta_{c1}$, the system exhibits Quasi-Long-Ranged Order (QLRO). 
(d)~Dependency of the scaling exponent $\beta$ on the density of obstacles at fixed noise strength $\eta=0.3$. The region where the system exhibits ordered phases is indicated by a vertical dashed line, where order is first LRO and then QLRO. 
System size $N_b=40000$ ($L=200$). }
\label{fig:gnf}
\end{figure}

\section{Anomalous density fluctuations and clustering statistics}
\label{sec:gnf}

A way to characterize the distribution of SPPs over the space is through the study of density fluctuations. 
Particularly useful information is provided by  the so-called number fluctuations. 
The idea is to divide the space over which particles move in cells of linear size $l$ and count the number of particles in each cell. Let us call $n(\mathbf{x}_i,l)$ the number of particles in the cell of linear size $l$ whose center is at position  $\mathbf{x}_i$. We are interested in computing the average of this quantity $\langle n(\mathbf{x}_i,l)  \rangle_i$ and its standard deviation $\sigma$. 
It can be shown~\cite{ramaswamy2010} that $\langle n(\mathbf{x}_i,l)  \rangle_i =\rho_b l^2=\langle n \rangle_l$ and that 
\begin{equation}
	\sigma (l) = \sqrt{ \langle \left(  n(x_i,l)-  \langle n \rangle_l)  \right)^2 \rangle_i } = \langle n \rangle_l^{\beta} \, ,
	\label{eq:gnf_definition}
\end{equation}
where the average $\langle \hdots \rangle_i$ is performed over the cells the space has been divided into.   
An exponent $\beta=1/2$ is expected for a random distribution or particles, while  $\beta>1/2$ corresponds to giant number fluctuations (GNF). 
It has been predicted that in spatially homogeneous systems, the SPPs in the ordered phase -- far away from the band regime -- exhibit GNF~\cite{ramaswamy2003}. 
This prediction has been observed in simulations of SPP particles, where a value of $\beta \sim 0.8$ was found~\cite{chate2008,gnf2012}. 
Here, we are interested in knowing the impact of an heterogeneous environment on the number fluctuations. 
Figure~\ref{fig:gnf} shows how number fluctuations are affected by changing the noise intensity $\eta$ for a fixed density of obstacles $\rho_o\approx 0.102$ (see Fig.~\ref{fig:gnf}(a),(c)) and by 
varying the obstacle density  $\rho_o$ while keeping the noise fixed, here $\eta=0.3$ (see Fig.~\ref{fig:gnf}(b),(d)). 
The top panels of Fig.~\ref{fig:gnf} correspond to the scaling of  $\sigma(l)$ with the average number $\langle n \rangle_l$ of particles per cell. In the figure, dashed blue lines correspond to a slope $1/2$, while red solid lines  to a slope $1$. 
Undoubtedly, GNF are also present in heterogeneous space. 
Nevertheless, there are important differences with what we know from SPP in homogeneous space. 
For instance, for a fixed (high enough) obstacle density $\rho_o$, GNF are suppressed, or at least decrease, as $\eta \to \eta_{c2}$, i.e. $\beta$ adopts smaller values as $\eta$ approaches $\eta_{c2}$, Fig.~\ref{fig:gnf}(c). 
We recall that in homogeneous space GNF are expected to be characterized by the same anomalous exponent as $\eta \to 0$. 
We also point out that in both homogeneous and heterogeneous space, number fluctuations become normal for $\eta>\eta_{c1}$, i.e. in high-noise disordered phase. 
This means that at high obstacle densities (i.e. for $\rho_o\geq 0.1$), GNF are stronger -- meaning that $\beta$ adopts its largest value -- at some point in between $\eta_{c2}<\eta_M<\eta_{c1}$, and this seems to occur close to $\eta_M$. 
On the other hand, if we fix the noise intensity $\eta$, we observe  that GNF  decay (i.e., $\beta$  goes down) as the density of obstacles $\rho_o$ is increased, reflecting the global tendency of the system to go to disorder when $\rho_o \to \infty$. 
We find that for $\rho_o \to 0$,  $\beta \to 0.8$ as expected in an homogeneous media, while as $\rho_o$ is increased, the exponent $\beta$ exhibits two regimes with $\rho_o$, approaching linearly for high obstacle densities $1/2$, where fluctuations can be considered normal and the system disordered, see  Fig.~\ref{fig:gnf}(d). 

\begin{figure}
\centering
\includegraphics[scale=0.35]{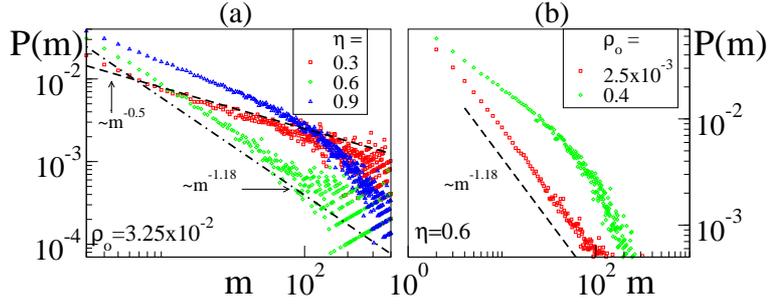}
\caption{Clustering statistics. (a) The cluster size distribution (CSD) $P(m)$ for the fixed density of obstacles $\rho_o=3.25\times 10^{-2}$ for different noises $\eta$. The critical noise values are  $\eta_{c2} \approx 0.05$ and $\eta_{c1} \approx 0.75$. The CSD for $\eta=0.9$ corresponds to a disordered phase, while the other two CSDs to ordered phases. The dashed and dotted-dashed curves provide the scaling $\propto m^{-1.18}$ and $\propto m^{-0.5}$ used as reference. (b) $P(m)$ for fixed noise value $\eta=0.6$  and two different values of the obstacle density  $\rho_o$. With $\rho_o=2.5 \times 10^{-3}$ the system is in an ordered phase and bands are observed, while with $\rho_o=0.4$ the system is fully disordered and bands are not observed.  The dashed line corresponds to the scaling $\propto m^{-1.18}$.   
System size $N_b=19600$ ($L = 140$).
}
\label{fig:clustering}
\end{figure}

Another alternative to study how particles are distributed in space is to look at the cluster size distribution. As before, we are interested in understanding how the presence of obstacles affects the non-equilibrium clustering statistics of the SPPs with respect to what we know from homogeneous media~\cite{peruani2006,peruani2010,peruani2013}.  
By ``cluster" we understand a group of connected particles, such that the distance between two connected particles is smaller or equal to the interaction radius.   
The size or mass of a cluster, which we denote here with the letter ``m", is the number of particles the cluster contains. 
Our quantity of interest is the (weighted) cluster size distribution (CSD) $P(m)$. Its definition is given by: 
\begin{eqnarray}
	P(m) = \lim_{t \to \infty} P(m,t) = \lim_{t \to \infty} \frac{m \, n_m(t)}{N} \, ,
	\label{eq:pm}
\end{eqnarray}
where $n_m(t)$ refers to the number of clusters of mass $m$ that are present in the system at time $t$. The limit is to indicate that we look at the steady state CSD and neglect transitory behaviors. 
Fig.~\ref{fig:clustering}(a) shows how the CSD is changed by varying the  noise $\eta$ for fixed $\rho_o=3.25\times 10^{-2}$. As a reference, the CSD corresponding to a fully disordered phase, i.e. $\eta=0.9$, is shown. 
We find that in between $\eta_{c1}$ and $\eta_M$ the CSD distribution is roughly power-law,  $ P(m) \propto m^{-\omega}$, with an exponent $\omega \sim1.18$ that falls in the  range $[0.8,4/3]$ as expected~\cite{peruani2013}. As we move to lower noise values, e.g. $\eta=0.3$, there is a strong depletion of isolated particles and small clusters and particles tend to form larger clusters, way larger than those observed close to  $\eta_{c1}$ or in the disordered phase (notice the log scale in the figure). 
Fig.~\ref{fig:clustering}(b) displays the CSD at fixed noise $\eta=0.6$ and two different values of $\rho_o$. 
For $\rho_o = 2.5\times 10^{-3}$ we observe bands and the CSD is again a power-law with exponent  $\omega \sim 1.18$, similar to what was reported for the 
Vicsek model~\cite{vicsek1995}  in the band regime where $\omega \sim 1.3$~\cite{peruani2013}.  
The figure evidences that  by increasing the density of obstacles $\rho_o$, at fixed noise, the functional form of CSD is dramatically affected. In particular, it shows that at very large obstacle density the CSD becomes exponential as expected for the disordered phase, with  a well defined average cluster size~\footnote{Power-law CSDs may be such that their first moment diverges, while exponential CSDs always have a well defined first moment.} and not surprisingly the system exhibits normal number fluctuations.

\section{Discussion: static vs dynamic heterogeneities and the symmetry of the interactions}
\label{sec:discussion}

\begin{figure}
\centering
\includegraphics[scale=0.35]{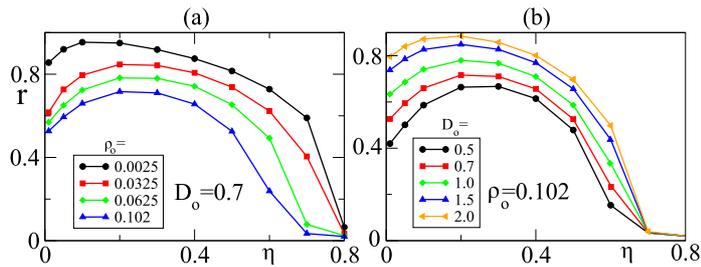}
\caption{SPP in a dynamical environment, where obstacle diffuse with a diffusion coefficient $D_o$. (a) Order parameter $r$ vs. $\eta$ for various obstacles densities $\rho_o$ and  constant diffusion coefficient $D_o$.  (b) $r$ vs. $\eta$ for constant density of obstacles $\rho_o$ and various diffusion coefficients $D_o$. Notice that there exists an optimal noise value even in a dynamical environment. System size $N_b=10000$ ($L=100$).}
\label{fig:different_cases}
\end{figure}

There is no reason to believe that static and dynamic heterogeneous environments lead to similar large-scale collective effects. 
In particular, the conclusions drawn from the finite size analysis performed with static obstacles, i.e. with a sort of ``quenched" noise, may not apply to dynamical heterogeneities. 
One may argue that dynamical heterogeneities may be mapped to an effective noise in a SPP system with homogeneous space. In this scenario, the dynamical heterogeneities should not affect 
qualitatively the large-scale  properties of the system but only have an impact on the critical point. 
A rigorous analysis would require a finite size study of SPPs in dynamical heterogeneous environments, which is  a very time-demanding numerical task  out of the scope of the current paper. 
Less ambitious but not less informative,  we can analyze the impact of a dynamical heterogeneous medium on the collective properties of SPPs in a fixed system size. 
Let us assume that the obstacles now diffusive over the space with a diffusion constant $D_o$. The position of the $k$-th obstacle obeys:
\begin{eqnarray}
 \dot{\mathbf{y}}_k = \sqrt{2 D_o} \xi_k(t) \,
 \end{eqnarray} 
 where $\langle \xi_k(t) \rangle = 0$ 
 and  $\langle \xi_k(t) \xi_f(t') \rangle = \delta(t-t') \delta_{k,f}$. 
Fig.~\ref{fig:different_cases} shows the order parameter $r$ as function of the (angular) noise $\eta$, for various values of $\rho_o$ and fixed obstacle diffusion coefficient $D_o=0.7$, panel (a), 
and fixed obstacle density $\rho_o=0.102$ and various obstacle diffusion coefficient $D_o$, panel (b). 
We find that even for a dynamical heterogeneous environment, there is an optimal noise that maximizes collective motion. As the obstacle density is increased, the level of ordering, i.e. $r$, decreases for all angular noises, Fig.~\ref{fig:different_cases}(a). 
On the other hand, we learn that the faster the obstacles diffuse, the weaker is the effect of the obstacles, Fig.~\ref{fig:different_cases}(b). Moreover, the numerical data suggests that in the limit of $D_o \to \infty$ the system behaves again as an homogeneous system with its critical point shifted to smaller noise values~\footnote{Compare  Fig.~\ref{fig:different_cases}(b) and Fig.~\ref{fig:order_parameter}(a), curve for $\rho_o=0$ to see the shift in the critical point.}. This suggests that in this limit effectively the problem can be mapped to an homogeneous system with an effective (angular) noise intensity. 

\begin{figure}
\centering
\includegraphics[scale=0.35]{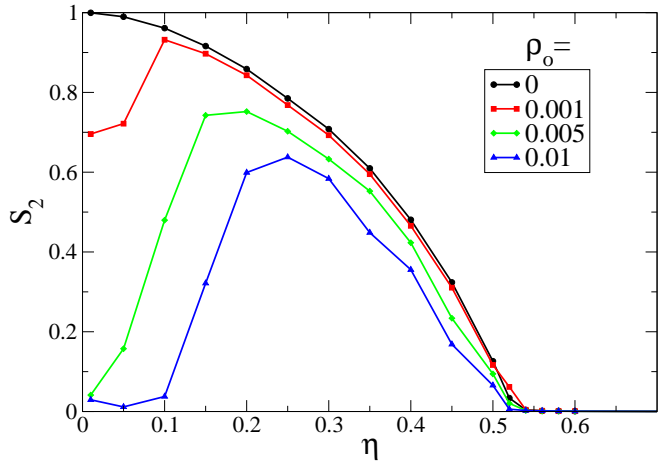}
\caption{SPPs with nematic alignment in a (static) heterogeneous medium. Nematic order parameter $S_2$ vs. noise intensity $\eta$ for different obstacle densities $\rho_o$. Notice that also SPPs with nematic alignment exhibit 
an optimal noise that maximizes order in the system. System size $N_b=10000$ ($L=100$).
}
\label{fig:nematic}
\end{figure}

Finally, we may wonder whether the observed optimal value is due to the particular symmetry of the velocity alignment between the SPPs that has been used, i.e. due to $q=1$ in Eq.~\ref{eq:direction}. 
To address this question, we perform simulations with SPPs interacting via a nematic velocity alignment, i.e. $q=2$ in Eq.~\ref{eq:direction}, that move in an heterogeneous medium with static obstacles. 
Since the alignment is nematic, we replace the order parameter by: $S_2 = \left\langle \left|
		\frac{1}{N_b}
		\sum_{i=1}^{N_b}
		e^{i \, 2\,\theta_i(t)}
		\right|
	\right\rangle_t$, where  $\langle \hdots \rangle_t$ represents a temporal average after a short transient. 
For a totally disordered system, $S_2=0$, while $S_2>0$ implies that the system exhibits, for the tested system size, (global) nematic order. 
%
Fig. 11 shows that optimal noise $\eta_M$ exists also for SPPs interacting via nematic alignment in an heterogeneous environment, as discussed above for ferromagnetic velocity alignment.
In this case, the optimal noise $\eta_M$ maximizes nematic ordering, i.e. it favors the emergence of a preferred direction of motion, where  50\% of the SPPs move roughly parallel to it and the other 50\% antiparallel to it. 
Details about SPPs with nematic velocity alignment in homogeneous media can be found in~\cite{peruani2008,ginelli2010,peshkov2012}.

\section{Conclusions}
\label{sec:conclusion}



%

We have learned that  even small levels of heterogeneity  lead to  qualitative changes in the large-scale properties of SPP systems interacting via a velocity alignment mechanism. 
Some  of  the new statistical features that emerge due to the presence of heterogeneities -- as for instance the existence of an optimal noise that maximizes order~\cite{ChepizhkoAltmannPeruani2013} -- are present in both statical and dynamical heterogeneous media, as well as by changing the symmetry of the velocity alignment mechanism of the SPPs.  
Other findings, as the absence of long-range order for high levels of heterogeneity (i.e. for $\rho_o\geq0.1$) apply exclusively to static obstacles. 
In general, we can conclude that in heterogeneous environments  the physics of SPP systems is different from what we know from homogeneous ones, 
with the presence of obstacles making more difficult for the SPPs to spread information about their moving direction across the system. 
From the two  information spreading mechanisms~\cite{meschede2013,toner2005} -- the one involving direct particle-particle interaction and responsible of 
order inside individual clusters, and the other one involving cluster-cluster information exchange, often through particle exchange among clusters --  
 obstacles affect the second one~\footnote{Information spreading in the context of active particles (without alignment) was studied with self-propelled disks that exchange their internal state upon collision~\cite{sibona2008}. As for aligning active particles, there are also two information mechanisms, with particle-particle information exchange being way faster than cluster-cluster information exchange. Interestingly, fluctuations play a major role even in the absence of cluster-cluster information exchange in two dimensions where the physics of the problem is always non-trivial~\cite{lee2013}.}.
In particular,  due to the obstacle presence, clusters get quickly uncorrelated as result of  independent collisions with the obstacles. 
At high obstacle densities or low noise amplitudes, particle exchange among clusters becomes insufficient 
to maintain the moving clusters correlated and the level of (global) order  decreases.   

The spatial arrangement of particles is also strongly affected by the presence of obstacles. 
The high-order traveling  bands -- reported to emerge in the classical, homogeneous, Vicsek model~\cite{gregoire2004,caussin2014} -- become less pronounced and even disappear as the level of 
heterogeneities, i.e. obstacles, is increased. 
At the point where bands are no longer observed, the ordering properties change from long-range to quasi-long range, which suggests that in the limit of an infinite system (keeping constant both, obstacle and particle density) and for static obstacles, the 
SPPs cannot maintain a coherent migratory route along  the (infinite) heterogeneous space: i.e. the system becomes disordered. 
On the other hand, our finite size analysis revealed that at high obstacle densities (i.e. $\rho_o\geq0.1$), the system exhibits two critical points, one at low and another one at high noise value. 
Finally, the study of density fluctuations  indicated that the giant-number-fluctuation exponent $\beta$ moves towards $1/2$, which corresponds to normal density fluctuations, as we approach the noiseless limit, as well as for large enough obstacle densities. 

Finally, it is worth mentioning that few experiments with active particle in heterogeneous media have been already performed, so far, with active particles without alignment: self-propelled janus particles moving on patterned surfaces~\cite{volpe2011} and speckle light fields~\cite{paoluzzi2014,volpe2014}. 
Interestingly, patterned regular environments have been initially used to rectify the motion of active swimmers such as bacteria in diluted suspensions 
 (i.e. in a non-interacting context)~\cite{galajda2007,wan2008,tailleur2009}. 
In these systems, volume exclusion effects and the size of the moving active particles play a central role. 
Such observations have triggered a good deal of theoretical work.  
For instance, in simulations with  self-propelled disks (SPD) it has been shown that SPDs can get locally jammed by the obstacles~\cite{reichhardt2014}~\footnote{Interestingly, SPD also exhibit  
an optimal noise that maximizes particle flux~\cite{reichhardt2014}. The explanation for this optimal noise seems to be rooted in the local jamming dynamics, being intrinsically different from the one related to the optimal noise value reported here.},   
while in simulations with SP rods it has been found that V-shaped obstacles can be used to trap particles~\cite{kaiser2012}. 
On the other hand, it has been shown in simulations with circularly moving active particles that a  
regular configuration of obstacles can be used to filter the active particles~\cite{mijalkov2013,reichhardt2013}, 
while narrow channels can direct particle motion~\cite{radtke2012}. 
It remains to be seen how the large-scale collective properties reported here are affected by introducing 
a velocity alignment mechanism in the above mentioned more realistic models. 

%
%
%
%
%

%
%
%


\bibliographystyle{iopart-num}

\begin{thebibliography}{10}
\expandafter\ifx\csname url\endcsname\relax
  \def\url#1{{\tt #1}}\fi
\expandafter\ifx\csname urlprefix\endcsname\relax\def\urlprefix{URL }\fi
\providecommand{\eprint}[2][]{\url{#2}}

\bibitem{schaller2010}
Schaller V, Weber C, Semmrich C, Frey E and Bausch A 2010 {\em Nature\/} {\bf
  467} 73

\bibitem{zhang2010}
Zhang H, Be{'}er A, Florin E~L and Swinney H 2010 {\em Proc. Natl. Acad. Sci.
  USA\/} {\bf 107} 13526

\bibitem{peruani2012}
Peruani F, Starruss J, Jakovljevic V, Sogaard-Andersen L, Deutsch A and B{\"a}r
  M 2012 {\em Phys. Rev. Lett.\/} {\bf 108} 098102

\bibitem{Buhl_science_locust}
Buhl J, Sumpter D~J~T, Couzin I~D, Hale J~J, Despland E, Miller E~R and Simpson
  S~J 2006 {\em Science\/} {\bf 312} 1402--1406

\bibitem{romanczuk2009}
Romanczuk P, Couzin I and Schimansky-Geier L 2009 {\em Phys. Rev. Lett.\/} {\bf
  102} 010602

\bibitem{Huepe_fish_2013}
{Tunstr{\o}m} K, Katz Y, Ioannou C~C, Huepe C, Lutz M~J and Couzin I~D 2013
  {\em PLoS Comput Biol\/} {\bf 9} e1002915

\bibitem{Holdo2011}
Holdo R~M, Fryxell J~M, Sinclair A~R~E, Dobson A and Holt R~D 2011 {\em PLoS
  ONE\/} {\bf 6} e16370

\bibitem{deseigne2010}
Deseigne J, Dauchot O and Chat{\'e} H 2010 {\em Phys. Rev. Lett.\/} {\bf 105}
  098001

\bibitem{kudrolli2008}
Kudrolli A, Lumay G, Volfson D and Tsimring L 2006 {\em Phys. Rev. E\/} {\bf
  74} 030904(R)

\bibitem{weber2013}
Weber C, Thueroff F and Frey E 2013 {\em arXiv:1301.7701\/}

\bibitem{jiang2010}
Jiang H~R, Yoshinaga N and Sano M 2010 {\em Phys. Rev. Lett.\/} {\bf 105}
  268302

\bibitem{golestanian2012}
Golestanian R 2012 {\em Phys. Rev. Lett.\/} {\bf 108} 038303

\bibitem{theurkauff2012}
Theurkauff I, Cottin-Bizzone C, Palacci J, Ybert C and Bocquet L 2012 {\em
  Phys. Rev. Lett.\/} {\bf 108} 268303

\bibitem{palacci2013}
Palacci J, Sacanna S, Steinberg A, Pine D and Chaikin P 2013 {\em Science\/}
  {\bf 339} 936

\bibitem{golestanian2009}
Golestanian R 2009 {\em Phys. Rev. Lett.\/} {\bf 102} 188305

\bibitem{paxton2004}
et~al W~P 2004 {\em J. Am. Chem. Soc.\/} {\bf 126} 13424

\bibitem{mano2005}
Mano N and Heller A 2005 {\em J. Am. Chem. Soc.\/} {\bf 127} 11574

\bibitem{rucker2007}
R{\"u}ckner G and Kapral R 2007 {\em Phys. Rev. Lett.\/} {\bf 98} 150603

\bibitem{howse2007}
Howse J, Jones R, Ryan A, Gough T, Vafabakhsh R and Golestanian R 2007 {\em
  Phys. Rev. Lett.\/} {\bf 99} 048102

\bibitem{golestanian2005}
Golestanian R, Liverpool T~B and Ajdari A 2005 {\em Phys. Rev. Lett.\/} {\bf
  94}(22) 220801

\bibitem{spp_activecolloid_nature2013}
Bricard A, Caussin J~B, Desreumaux N, Dauchot O and Bartolo D 2013 {\em
  Nature\/} {\bf 503} 95--98 ISSN 0028-0836

\bibitem{peruani2007}
Peruani F and Morelli L 2007 {\em Phys. Rev. Lett.\/} {\bf 99} 010602

\bibitem{romanczuk2012}
Romanczuk P, B{\"a}r M~M, Ebeling W, Lindner B and Schimansky-Geier L 2012 {\em
  Eur. Phys. J. Special Topics\/} {\bf 202} 1


\bibitem{dworkin}
Dworkin M 1993 {\em Myxobacteria II\/} (Amer Society for Microbiology)

\bibitem{marchetti2013}
Marchetti M~C, Joanny J~F, Ramaswamy S, Liverpool T~B, Prost J, Rao M and Simha
  R~A 2013 {\em Rev. Mod. Phys.\/} {\bf 85} 1143--1189

\bibitem{ramaswamy2010}
Ramaswamy S 2010 {\em Annual Review of Condensed Matter Physics\/} {\bf 1}
  323--345

\bibitem{vicsek2012}
Vicsek T and Zafeiris A 2012 {\em Physics Reports\/} {\bf 517} 71--140

\bibitem{ramaswamy2003}
S~Ramaswamy R~A~S and Toner J 2003 {\em Europhys. Lett.\/} {\bf 62} 196--202

\bibitem{vicsek1995}
Vicsek T, A~Czirok E, Jacob E~B, Cohen I and Shochet O 1995 {\em Phys. Rev.
  Lett.\/} {\bf 75} 1226

\bibitem{gregoire2004}
Gr{\'e}goire G and Chat{\'e} H 2004 {\em Phys. Rev. Lett.\/} {\bf 92} 025702

\bibitem{caussin2014}
Caussin J~B, Solon A, Peshkov A, Chat\'e H, Dauxois T, Tailleur J, Vitelli V
  and Bartolo D 2014 {\em Phys. Rev. Lett.\/} {\bf 112} 148102

\bibitem{spp_tonertu_prl1995}
Toner J and Tu Y 1995 {\em Physical Review Letters\/} {\bf 75} 4326--4329

\bibitem{toner1998}
Toner J and Tu Y 1998 {\em Phys. Rev. E\/} {\bf 58} 4828

\bibitem{ChepizhkoAltmannPeruani2013}
Chepizhko O, Altmann E~G and Peruani F 2013 {\em Phys. Rev. Lett.\/} {\bf
  110}(23) 238101

\bibitem{peruani2008}
Peruani F, Deutsch A and B{\"a}r M 2008 {\em Eur. Phys. J. Special Topics\/}
  {\bf 157} 111

\bibitem{ChepizhkoPeruani2013}
Chepizhko O and Peruani F 2013 {\em Phys. Rev. Lett.\/} {\bf 111}(16) 160604

\bibitem{Binder}
Binder K 1997 {\em Reports on Progress in Physics\/} {\bf 60} 487

\bibitem{chate2008}
Chat{\'e} H, Ginelli F, Gr{\'e}goire G and Raynaud F 2008 {\em Phys. Rev. E\/}
  {\bf 77} 046113

\bibitem{Kosterlitz}
Kosterlitz J~M and Thouless D~J 1973 {\em Journal of Physics C: Solid State
  Physics\/} {\bf 6} 1181

\bibitem{gnf2012}
Dey S, Das D and Rajesh R 2012 {\em Phys. Rev. Lett.\/} {\bf 108}(23) 238001

\bibitem{peruani2006}
Peruani F, Deutsch A and B{\"a}r M 2006 {\em Phys. Rev. E\/} {\bf 74} 030904(R)

\bibitem{peruani2010}
Peruani F, Schimansky-Geier L and B{\"a}r M 2010 {\em Eur. Phys. J. Special
  Topics\/} {\bf 191} 173--185

\bibitem{peruani2013}
Peruani F and B{\"a}r M 2013 {\em New J. Phys.\/} {\bf 15} 065009

\bibitem{ginelli2010}
Ginelli F, Peruani F, B{\"a}r M and Chat{\'e} H 2010 {\em Phys. Rev. Lett.\/}
  {\bf 104} 184502

\bibitem{peshkov2012}
Peshkov A, Aranson I, Bertin E, Chate H and Ginelli F 2012 {\em Phys. Rev.
  Lett.\/} {\bf 109} 268701

\bibitem{volpe2011}
Volpe G, Buttinoni I, Vogt D, Kummerer H~J and Bechinger C 2011 {\em Soft
  Matter\/} {\bf 7} 8810--8815

\bibitem{paoluzzi2014}
Paoluzzi M, Leonardo R~D and Angelani L 2014 {\em Journal of Physics: Condensed
  Matter\/} {\bf 26} 375101

\bibitem{meschede2013}
Meschede M and Hallatschek O 2013 {\em New Journal of Physics\/} {\bf 15} 4

\bibitem{toner2005}
Toner J and Tu Y and Ramaswamy S 2005 {\em Ann. Phys.} {\bf 318} 170-244

\bibitem{sibona2008}
Peruani F and Sibona G 2008  {\em Phys. Rev. Lett.} {\bf 100} 168103 

\bibitem{lee2013}
Peruani F and Lee CF 2013,  {\em Europhys. Lett.} {\bf 102} 58001

\bibitem{volpe2014}
Volpe G, Volpe G and Gigan S 2014 {\em Scientific Reports\/} {\bf 4} 3936

\bibitem{galajda2007}
Galajda P, Keymer J, Chaikin P and Austin R 2007 {\em J. Bacterial.\/} {\bf
  189} 8704

\bibitem{wan2008}
Wan M, Reichhardt C~O, Nussinov Z and Reichhardt C 2008 {\em Phys. Rev.
  Lett.\/} {\bf 101} 018102

\bibitem{tailleur2009}
Tailleur J and Cates M 2009 {\em Europhys. Lett.\/} {\bf 86} 60002

\bibitem{reichhardt2014}
Reichhardt C and Olson~Reichhardt C~J 2014 {\em Phys. Rev. E\/} {\bf 90}(1)
  012701

\bibitem{kaiser2012}
Kaiser A, Wensink H and L{\"o}wen H 2012 {\em Phys. Rev. Lett.\/} {\bf 108}
  268307

\bibitem{mijalkov2013}
Mijalkov M and Volpe G 2013 {\em Soft Matter\/} {\bf 9} 6376--6381

\bibitem{reichhardt2013}
Reichhardt C and Reichhardt C~J~O 2013 {\em Phys. Rev. E\/} {\bf 88}(4) 042306

\bibitem{radtke2012}
Radtke P and Schimansky-Geier L 2012 {\em Phys. Rev. E\/} {\bf 85} 051110


\end{thebibliography}


 








\end{document}